# GoS Proposal to Improve Trust and Delay of MPLS Flows for MCN Services

Francisco J. Rodríguez-Pérez
Computer Science Dept., Area of Telematics Engineering
University of Extremadura
Cáceres, Spain

José-Luis González-Sánchez
Computer Science Dept., Area of Telematics Engineering
University of Extremadura
Cáceres, Spain

Alfonso Gazo-Cervero
Computer Science Dept., Area of Telematics Engineering
University of Extremadura
Cáceres, Spain

*Abstract*—In this article, Guarantee of Service (GoS) is defined as a proposal to improve the integration of Mission Critical Networking (MCN) services in the Internet, analyzing the congestion impact on those privileged flows with high requirements of trust and delay. Multiprotocol Label Switching (MPLS) is a technology that offers flow differentiation and QoS in the Internet. Therefore, in order to improve network performance in case of congested domains, GoS is proposed as a technique that allows the local recovering of lost packets of MPLS privileged flows. To fulfill the GoS requirements for integration of MCN in MPLS, a minimum set of extensions to RSVP-TE has been proposed to provide GoS capable routes. Moreover, we have carried out an analytical study of GoS scalability and a performance improvement analysis by means of simulations.

## I. INTRODUCTION

The integration of Mission Critical Networking (MCN) with the Internet allows enhancing reachability and ubiquity and the cost reduction of deployment and maintenance. However, an efficient network operation for MCN services is always required, but the Internet is a heterogeneous network that typically includes numerous resource-constrained devices [1], which creates bottlenecks that affect the network performance. In this context, Multiprotocol Label Switching (MPLS) is currently used to provide policy management for heterogeneous networks and protocols with QoS integration purposes, combining traffic engineering capabilities with flexibility of IP and class-of-service differentiation [2], [3].

MPLS Label Switched Paths (LSP) let the head-end Label Edge Router (LER) to control the path that traffic takes to a particular destination [4]. This method is more flexible than forwarding traffic based on destination address only. LSP tunnels also allow the implementation of a variety of policies related to the optimization of network performance [5]. Moreover, resilience allows LSP tunnels being automatically routed away from network failures or congestion points [6], [7]. Resource Reservation Protocol with Traffic Engineering (RSVP-TE) is the signalling protocol used to allocate resources for those LSP tunnels across the network [8]. Therefore, MPLS allocates bandwidth on the network when it uses RSVP-TE to build LSP [9]. When RSVP-TE is used to allocate bandwidth for a particular LSP, then the concept of *consumable resource* in the network is introduced, in order to allow edge nodes finding paths across the domain, which has bandwidth available to be allocated. However, there is no forwarding-plane enforcement of a reservation, which is signalled in the control plane only, which means that, for instance, if a Label Switch Router (LSR) makes a RSVP-TE reservation for 10 Mbps and later it needs 100 Mbps, it will congest that LSP [10]. The network attempts to deliver the 100 Mbps, causing a lower performance to other flows that can have even more priority, unless we attempt to apply traffic policing using QoS techniques [11]. In this context, extensions of RSVP-TE protocol are expected to be an important application for performance improvement in such problematic instances, because MPLS-TE is providing fast networks, but with no local flow control. Therefore, it is being assumed that devices are not going to be congested and that they will not lose traffic. However, resource failures and unexpected congestions cause traffic looses [12], [13]. In these cases, upper layers protocols will request re-transmissions of lost data at end points [14], [15], but the time interval to obtain re-transmitted data can be significant for some types of time-critical MCN applications, such as real-time data delivery or synchronized healthcare services, where there are time-deadlines to be met.

The objective of this work is to analyze our Guarantee of Service (GoS) proposal as a resource engineering technique for local recovery of lost packets of MCN services, which need reliable and timely responses. With this purpose, GoS extensions of RSVP-TE [16] are used as a service-oriented technique, offering Privileged LSP to mission critical flows, in order to manage high requirements of delay and reliability. Furthermore, GoS does not propose the replacement of nodes in a MPLS domain but the incorporation of several GoS

This work is supported in part by the Regional Government of Extremadura (Economy, Commerce and Innovation Council) under GRANT PDT07A039.

capable MPLS nodes in bottlenecks. This way, in case of MCN services packets loss in a congested node, there will be a set of upstream nodes to request a local re-transmission to, increasing possibilities of finding lost packets faster.

The remainder of this article is structured as follows: Firstly, in Section 2, we define the GoS concept to be applied to MPLS flows for MCN services and how to signal the local recovery messages. Then, in Section 3 the proposed RSVP-TE extensions are studied, with the aim of minimizing the forwarding of GoS information across the MPLS domain. Next, an analysis of the GoS scalability is shown in fourth Section. In Section five, end-to-end (E-E) and GoS recoveries performances are compared by means of simulations [17], [18]. Finally we draw up some conclusions, results and contributions of our research.

## II. GUARANTEE OF SERVICE IN AN MPLS DOMAIN

Our GoS technique can be defined as the possibility for resilience improvement in congested networks to flows with high requirements of delay and reliability. In particular, the GoS for MPLS protocol provide LSR nodes with the capacity to recover locally lost packets of a MPLS flow for MCN services. The GoS proposal is provided by a limited RSVP-TE protocol extension, to achieve GoS capacity in intermediate nodes, in order to get faster re-transmissions of lost packets. Furthermore, our proposal let RSVP-TE to get local recoveries in case of LSP failures by means of *Fast Reroute* point-to-point technique. In [6] the efficiency of this technique was studied and compared to other E-E failure recoveries techniques.

Therefore, a buffer in GoS nodes to temporally store only packets of a MCN service is needed. However, a particular packet is only needed to be buffered for a short interval of time. This is because the time for a local recovery request for such packet to be received is very limited due to the low packets delay in MPLS backbones. So, a GoS node only needs to store a limited number of packets per flow, allowing very efficient buffer searches. This set of GoS nodes, which have switched the packets of a GoS flow, is called *GoS Plane* (*GoSP*) and the number of necessary hops to achieve a successfully local recovery is called *Diameter* (*d*) of the local re-transmission. This way, a greater GoS level gives a higher probability to achieve a local retransmission with lower diameter. Therefore, the diameter is the key parameter of a GoS re-transmission. In this paper we focus on an analysis of the diameter scalability.

In Fig. 1, operation of GoS is shown when a packet of a MCN service is discarded, for instance, in intermediate node $X_4$ and three feasible diameters can be used to recover locally the lost packet.

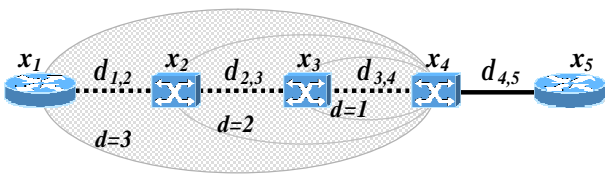

Figure 1. GoSP from node $X_4$, with diameter = 3 hops

GoS characterization information of a MCN flow packet consists of *GoSP, GoS Level* and *Packet ID*. GoSP is the most generic information. It is a constant value for every packet of flows in a same LSP. Therefore, it is related to the LSP, but neither to flows nor to packets. *GoS Level* is a constant value for every packet of a flow; i. e., it is flow specific information. A greater GoS level implies a greater probability that a packet can be re-transmitted from a previous hop, because a flow with a higher GoS level is signalled across an LSP with more GoS capable nodes. Moreover, more memory is allocated in GoS buffers for flows with the highest GoS level. It allows classifying the GoS priority level with respect to other MCN flows of the LSP or other paths in the domain. Moreover, this value keeps constant only in packets belonging to the same MPLS Forwarding Equivalence Class (FEC). Finally, *Packet ID* is necessary to request local re-transmissions in case of packet loss of a MCN service. It is packet specific information, with a unique value per packet of a flow.

In order to get the GoSP from a GoS node when a MCN flow packet is lost, we consider a domain $G(U)$, with a set of nodes $U$ and a data flow $j(G)=j(x_i, x_n)$ in $G(U)$ across a path $LSP_{i,n}$, with the origin in node $x_i$ and destination in node $x_n$, with $\{x_i, x_n\} \subset U$. Maybe $x_n$ only knows incoming port and incoming label of any arrived packet of flow $j(G)$, i.e., $x_n$ only knows that $x_{n-1}$ is the sender of $j(x_i, x_n)$. It would know which node the sender of a packet is, using label information. However, this is not a reliable strategy because, in case of flow aggregates, an RSVP-TE aggregator could perform reservation aggregation to merge $k$ flows, in the form:

$$j(x_{n-1}, x_n) = \sum_{i=1}^{k} j_i(x_{n-1}, x_n) \quad (1)$$

Furthermore, $x_n$, may not be able to satisfy the Flow Conservation Law due to congestion:

$$\sum_{i=1}^{k} p_{il} > \sum_{j=1}^{k} p_{lj} \quad (2)$$

The parameter $p_{ij}$ is the traffic volume sent from $x_i$ to $x_j$ across $x_l$. Therefore, one or more packets are being discarded in $x_l$, because the number of outgoing packets from $x_l$ is lower than the number of incoming packets. In this case upper layers protocols will have to detect lost packets and re-transmit them from head-end.

In order to request local re-transmissions when a packet of a MCN service is lost, it is necessary for GoS to know the set of nodes that forward the GoS packets. Thus, $x_n$ would know that discarded traffic have been stored in the upstream GoS nodes of $LSP_{i,n}$. The first node to request a local re-transmission is the previous GoS capable neighbour. With this purpose, RSVP-TE has been extended to allow signalling the GoS re-transmission requests, even, across non-GoS nodes. This proposal avoids the re-transmissions requests to the head-end and brings a lesser increment of global $j(G)$ in the congested domain. Moreover, the deployment of GoS does not imply the replacement of a lot of routers in a MPLS domain, but only the insertion of several GoS capable nodes in bottlenecks. For this purpose, a study of distribution of GoS nodes in the domain has been carried out in order to get the optimal placement of GoS nodes. It has been carried out basing on several parameters, such as domain topology, links capacity, RSVP-TE reservations, network load and GoS level of the flows. The main benefit of this study is to minimize the diameter of local recoveries in case of MCN service data loss.

*A. A Connection-Oriented GoSP*

The throughput of a flow could be lower if GoS characterization information was carried with data packets. To avoid this, GoS information carried into data packets has been minimized, signalling the GoSP when the LSP is being signalled by RSVP-TE. This task is only carried out at the beginning, before data packets forwarding. Therefore, a GoS integrated with the MPLS Control Plane (CP), avoids that GoS information must be forwarded with every MPLS data packet. This way, GoS characterization info (GoS Level and GoSP previous hop) is only sent when LSP is being signalled, adding a new row in a table of the GoS nodes. This is similar to the operation of RSVP-TE protocol when an LSP is signalled across the domain, considering the GoSP as a connection-oriented subset of nodes of the LSP with GoS capability. The LSP that supports a GoSP to forward a MCN service with high requirements of delay and reliability is named *privileged LSP*.

This way, GoS proposal extends the RSVP-TE protocol to let GoSP signalling as a subset of nodes of a privileged LSP. In the CP, when a node receives an RSVP-TE message requesting a new LSP, it inserts a new row in the *Forwarding Information Base* (FIB), about how to forward data packets across nodes of the LSP that is being signalled. Therefore, this is the info to be used by an LSR in the MPLS Forwarding Plane (FP) when it receives a MPLS packet to be switched. With FIB information it will know how to make the label swapping and how to forward it to the next hop. Therefore, with a connection-oriented GoSP, a GoS node that in FP detects an erroneous or discarded privileged packet, it only needs to get the FEC and GoS packet ID of the lost packet, because the GoS table already has all it needs to initiate a local re-transmission request. When RSVP-TE signals a new LSP for a MCN flow, then every GoS capable node of the LSP will add a new row to the FIB table, but also to the GoS Table. Flows information in that table is very simple, as in Table 1 is shown.

TABLE I. AN EXAMPLE OF GOS TABLE VALUES

| FEC | GoS Level | GoSP PHOP |
|---|---|---|
| 35 | 0000000000001011 | x.x.160.12 |
| 36 | 0000000000000001 | x.x.160.73 |
| 37 | 0000000000010010 | x.x.160.17 |
| 38 | 0000000000000001 | x.x.160.35 |

The table includes a first column for FEC or flow identification, a second column for flow GoS level and, finally, a third column is used to know the previous GoS hop address, to send it a request in case of GoS packet loss.

*B. Guarantee of Service States Diagram*

In Fig. 2 a states diagram of the operation of a GoS node is shown. In the FP, the state of a GoS node is *Data Forwarding*, switching labels and forwarding data packets to the next node. There are only two events that change this state in the GoS node. The first event is the detection of a GoS packet loss. In this case, the GoS capable node gets FEC and GoS packet identification and change its state to *Local recovery request*, sending a local re-transmission request (*GoSReq*) to the first node of GoSP (the closest upstream GoS node). When a response (*GoSAck*) is received, it changes to the initial state.

The other event that changes the state is reception of a *GoSReq* from any downstream GoS node, which is requesting a local re-transmission. In this case, the node changes its state to *Buffer Access*, to search the requested packet according to the information received in the *GoSReq*. If the requested packet is found in the GoS buffer, a *GoSAck* is sent in response to the *GoSReq*, indicating that requested packet was found and it will be re-transmitted locally. Therefore, it changes to *Local Re-transmission* state to get the GoS packet from the GoS buffer and re-forward it. Next, it will return to initial *Forwarding* state. In case of not find the packet in GoS buffer, it will send a *GoSAck* message, indicating that packet was not found and changing to *Local Recovery Request* state, sending a new *GoSReq* to its previous GoS node in the GoSP, if it is not the last one.

III. GUARANTEE OF SERVICE MESSAGES

GoS levels can easily be mapped to MPLS FEC, which is commonly used to describe a packet-destination mapping. A FEC is a set of packets to be forwarded in the same way (e.g. using the same path or Quality of Service criteria). One of the reasons to use the FEC is that allows grouping packets in classes. It can be used for packet routing or for efficient QoS supporting too; for instance, a high priority FEC can be mapped to a healthcare service or a low priority FEC to a web service.

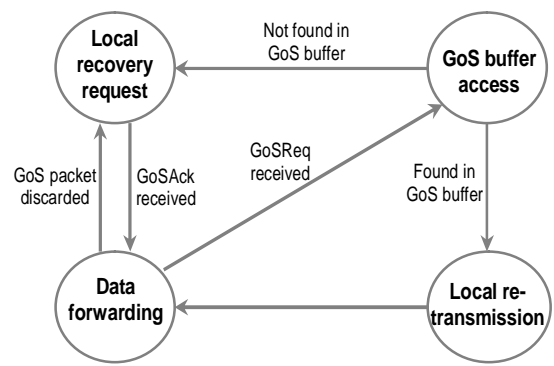

Figure 2. States diagram of a GoS capable node

Label is used by MPLS to establish the mapping between FEC and packet, because an incoming and outgoing labels combination identifies a particular FEC. With different classes of services, different FEC with mapped labels will be used. In our proposal, *GoS FEC* concept is used to classify the different GoS levels, giving more priority to the most privileged FEC. Therefore, GoS FEC will allow giving different treatments to GoS packets belonging to flows with different privileges, although they are being forwarded along the same path. With the purpose of minimize GoS signalling in the MPLS FP, GoS characterization info (GoS Level, Packet Id and GoSP) can be signalled by RSVP-TE in the MPLS CP. When a privileged LSP is being established, extended RSVP-TE *Path* and *Resv* messages can forward GoS Level and GoSP info (see Figs. 3 and 4).

When an LSP tunnel is being signalled in the CP, a GoS node that receives a GoS-extended Path message will access this GoS info to update its GoS Table. Then, it will record its IP address in the *GoSP PHOP* field of the *GoSPath* object because it will be the previous hop of the next downstream GoS node that detects a packet loss. It is not necessary to transport the entire GoSP in the *GoSPath* message, but only the last GoS node, because the node that detects a packet lost only send a local retransmission request to the PHOP in the GoSP. If PHOP cannot find the requested packet, it will request a local retransmission to the GoS PPHOP of the point of loss (if it is not the last one). Finally, following the RSVP-TE operation way, when an LSP is being signalled, GoS information will be confirmed with the reception of a GoS-extended *Resv* message, confirming the requested GoS level.

*A. Signalling of GoS Local Re-transmissions*

It is not necessary to send GoSP in every *GoSReq* message, because GoS nodes have an entry in the GoS Table with the GoSP PHOP to every flow. Therefore, in case that a GoSP PHOP node cannot satisfy a local re-transmission request, then it will get the GoS PHOP from the GoS Table, to send a new GoSReq to its GoSP PHOP to forward the request. So, it is not necessary that a node, which initiates a GoSReq, sends more requests to previous nodes of the GoSP PHOP. This technique has benefits in the LSP overhead when sending GoSReq messages. This is the reason to only buffer one address in the GoSP PHOP column, instead of the entire GoSP.

Therefore, in case of packet loss in a GoS node, this LSR would send to the upstream GoS PHOP a local re-transmission request. With this purpose, RSVP-TE Hello message has been extended. In particular, *Hello Request* message (see Fig. 5) has been extended with a *GoSReq* object, in order to allow requesting to the upstream GoSP PHOP the re-transmission of the lost packet specified in *Packet ID* field of the flow (specified in *Privileged Flow ID* field). Upstream GoS node that receives the *GoSReq* message sends a response in an extended *Hello Ack* message (see Fig. 6), with a *GoSAck* object to notify if requested packet has been found in the GoS buffer. Furthermore, following the RSVP-TE operation way, *Source Instance* and *Destination Instance* of the *Hello object* are used to test connectivity between GoSP neighbour nodes.

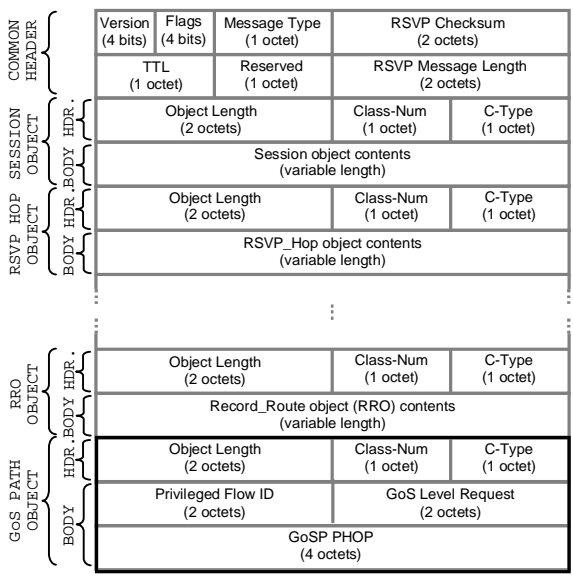

Figure 3. GoS extended Path message format with GoS Path object

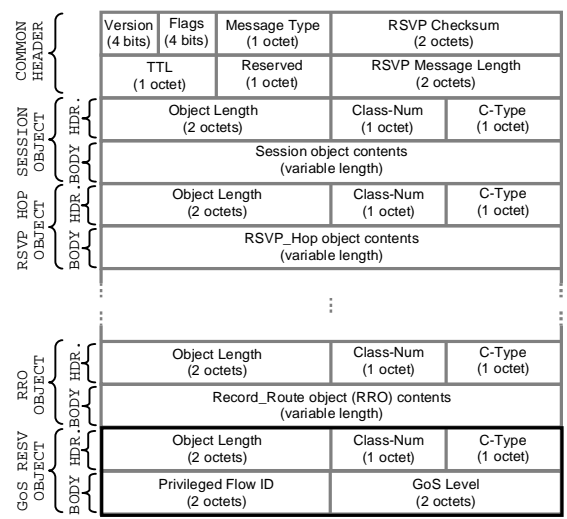

Figure 4. GoS extended Resv message format with GoS Resv object

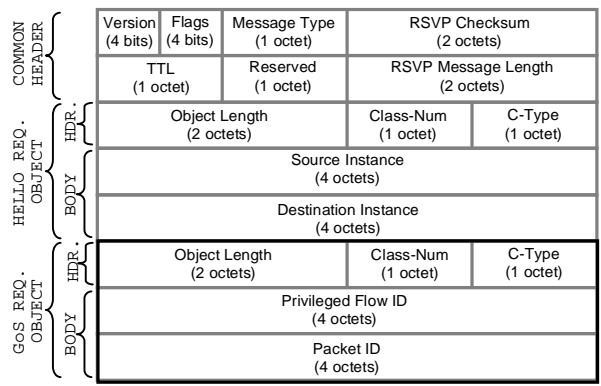

Figure 5. GoS extended Hello message format, with GoS Request object after the Hello object

IV. SCALABILITY OF THE GoSP DIAMETER

In this section we analyze the scalability of the connection-oriented GoSP. A MPLS domain $G(U)$ will be considered, with a set $X$ of $n$ nodes and a set $U$ of links. Let $d_{ij}$ the delay of link $(x_i, x_j) \in U$ and let $d(x_i, x_j)$ the delay of a path between two any nodes $x_i$ and $x_j$. Finally, let $d_{GoS}$ the delay proportion used for transmission of GoS characterization information in FP (*GoS packet ID*). The main objective is to analyze the scalability of the GoSP when lost packets are re-transmitted between two any nodes of $LSP_{i,n}$ in $U(G)$. This way, minimum delay used by a packet when is forwarded between two nodes of the path $LSP_{i,n}$ of $G(U)$ is:

$$\min d(x_i, x_j) = \sum_{i=1}^{n} \sum_{j=1}^{n} d_{ij} \, x_{ij} \qquad (3)$$

subject to:

$$\sum_{l=2}^{n} x_{1l} = 1 \qquad (4)$$

$$\sum_{i=1}^{n} x_{il} - \sum_{j=1}^{n} x_{lj} = 0, \; l = 2, 3, ..., n-1 \qquad (5)$$

$$\sum_{l=1}^{n-1} x_{ln} = 1 \qquad (6)$$

where: $x_{i,j} = 1, \forall (x_i, x_j) \in LSP_{i,n}$,
$x_{i,j} = 0, \forall (x_i, x_j) \notin LSP_{i,n}$ and $d_{i,i} = 0, \forall i$

A. *End-to-End Retransmissions*

Let $x_n$ a non-GoS congested end node. In case of packet discarding by $x_n$, then *Discarding Detection Time* ($DDT_{E-E}$) function between two nodes of $LSP_{i,n}$ is:

$$DDT_{E-E}(x_i, x_n) = \sum_{l=i}^{n-1} d_{l,l+1} \, x_{l,l+1} \qquad (7)$$

Minimal delay of the end to end (*E-E*) retransmission is:

$$d_{E-E}(x_i, x_n) = 2\sum_{l=i}^{n-1} d_{l,l+1} \, x_{l,l+1} \qquad (8)$$

Therefore, total delay $\Delta_{E-E}(x_i, x_n)$ to get discarded flow in $x_n$ is got from Eqs. (7) and (8):

$$\Delta_{E-E}(x_i, x_n) = 3\sum_{l=i}^{n-1} d_{l,l+1} \, x_{l,l+1} \qquad (9)$$

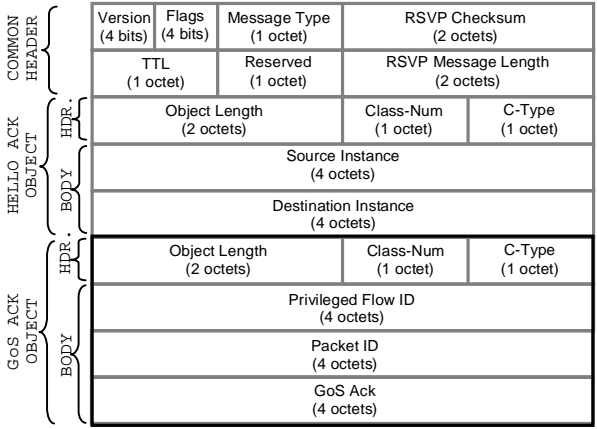

Figure 6. GoS extended Hello message format, with GoS Ack object after the Hello object

In Fig. 7, operation of the GoS when a packet that is being forwarded from $X_1$ to $X_5$ (with delay $d_{1,5}$) is discarded in the intermediate node $X_4$ is shown. For instance, in this case 3 GoSP diameters ($d$=1, $d$=2 and $d$=3) can be used to achieve a successfully local re-transmission. First, $X_4$ sends a local re-transmission request (*GoS_Req*) to the first node of the GoSP ($X_3$). Then, that node will send a response (*GoS_Ack*) to indicate whether it has found the requested packet or not in the GoS buffer. If it is found ($d$=1), it will send that locally recovered packet (*LRP*) towards its destination. But if it is not found, $X_3$ will send a new *GoS_Req* message to its PHOP in the GoSP ($X_2$). If $X_2$ finds requested packet, the successfully diameter would be $d$=2. Finally, if $X_1$, which is the last node of the GoSP, finds the lost MCN packet, then a diameter $d$=3 would achieve a successfully local re-transmission. Furthermore, this local recovery process is compared with both end-to-end re-transmission request (EERR) and end-to-end re-transmission packet (EERP).

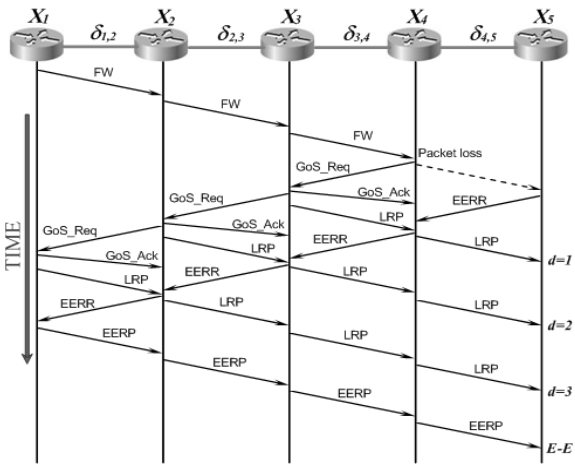

Figure 7. Local re-transmission operation when a GoS packet is discarded in an intermediate node

*B. GoS-based Local Re-transmissions*

Let $x_n$ be a GoS congested end node. In case of packet discarding by $x_n$, then *Discarding Detection Time* ($DDT_d$) between source and sink nodes of path $LSP_{i,n}$ is:

$$DDT_d(x_i, x_n) = \sum_{l=i}^{n-1} d_{l,l+1} \cdot d_{GoS} \cdot x_{l,l+1} \quad (10)$$

Minimal delay of local retransmission using a GoSP with diameter $d$ ($d_d$) is:

$$d_d(x_i, x_n) = 2\sum_{l=n-d}^{n-1} d_{l,l+1} \cdot d_{GoS} \cdot x_{l,l+1} \quad (11)$$

subject to: $0 < d < n - i$

If the diameter in Eq. (11) was $n-i$, then if $l = n-d = n-(n-i) = n-n+i = i$, we get that:

$$2\sum_{l=n-d}^{n-1} d_{l,l+1} \cdot d_{GoS} \cdot x_{l,l+1} = 2\sum_{l=i}^{n-1} d_{l,l+1} \cdot d_{GoS} \cdot x_{l,l+1} \quad (12)$$

i.e., it would be an *E-E* retransmission.

Moreover, if in Eq. (11) GoSP diameter was bigger than $n-i$, then it would be trying to get a retransmission from a previous node to $x_i$, but this one is the source of data flow, so it is unfeasible. Thus, total delay $\Delta_d(x_i, x_n)$ to get discarded traffic from initial instant of transmission is got from Eqs. (10) and (11):

$$\Delta_d(x_i, x_n) = \sum_{l=i}^{n-1} d_{l,l+1} d_{GoS} x_{l,l+1} + 2\sum_{l=n-d}^{n-1} d_{l,l+1} d_{GoS} x_{l,l+1} \quad (13)$$

At this point we test if Eq. (13) < Eq. (9):

$$\sum_{l=i}^{n-1} d_{l,l+1} d_{GoS} x_{l,l+1} + 2\sum_{l=n-d}^{n-1} d_{l,l+1} d_{GoS} x_{l,l+1} < 3\sum_{l=i}^{n-1} d_{l,l+1} d_{GoS} x_{l,l+1} \quad (14)$$

$$3\sum_{l=i}^{n-1} d_{l,l+1} x_{l,l+1} > d_{GoS} \sum_{l=i}^{n-1} d_{l,l+1} x_{l,l+1} + 2 d_{GoS} \sum_{l=n-d}^{n-1} d_{l,l+1} x_{l,l+1} \quad (15)$$

$$\sum_{l=i}^{n-1} d_{l,l+1} x_{l,l+1} > \frac{\left(2 d_{GoS} \sum_{l=n-d}^{n-1} d_{l,l+1} x_{l,l+1}\right)}{(3 - d_{GoS})} \quad (16)$$

In Eq. (16) the half-plane of solutions has been obtained for the case of a local recovery with diameter $d$ that have lower delay than an E-E re-transmission. Therefore, to get the GoSP diameter scalability with respect to the number of nodes of the privileged LSP and $\delta_{GoS}$, we get parameter $d$:

$$d < ((n-1-i) \cdot (3 - d_{GoS}))/(2 \cdot d_{GoS}) + 1 \quad (17)$$

In Fig. 8 scalability of the GoSP diameter for different LSP sizes (parameters $i$ and $n$) is shown. In chart we can see that there is a lineal rise when increasing the number of nodes of the LSP, until a maximum LSP size of 251 nodes. After this point, the maximum feasible diameter that would allow a successfully local re-transmission has a value of 250 hops.

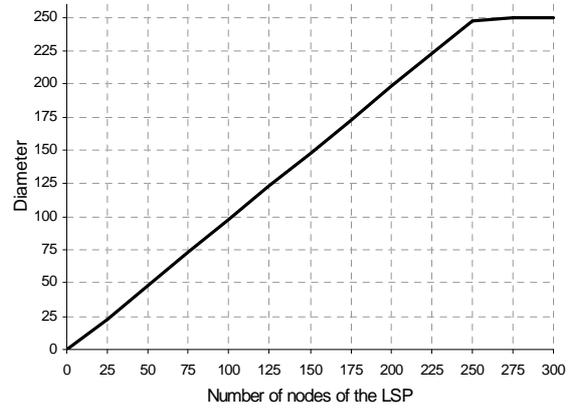

Figure 8. Scalability of GoSP diameter for different LSP sizes

This proof can easily be extended to include the case where an intermediate node $X_{DD}$ is requesting re-transmission, getting the same half-plane of solutions for the GoSP diameter, as is shown in Eq (17).

V. SIMULATION RESULTS

In order to evaluate the performance of GoS approach, we have carried out a series of simulations focused on AT&T backbone network topology (see Fig. 9), which is MPLS enabled to provide QoS for customers who require value-added services. In our simulations, AT&T core topology is characterized by 120 LER nodes, 30 LSR nodes and 180 links, with capacities in the range of [45Mbps, 2.5Gbps]. A GoS enabled node has been located at the eight routers with the biggest connectivity. In scenarios, signalled LSP are uni-directional and the bandwidth demanded for each flow is drawn from a distribution over the range of [64Kbps, 4Mbps]. In order to analyze the effect that GoS re-transmissions have on transport layer protocols, several MCN services over TCP/IP that use LSP across a different number of GoS capable nodes have been compared with not privileged TCP/IP flows across the same paths. LSP congestion has also been considered in the range of [0.01%, 4%].

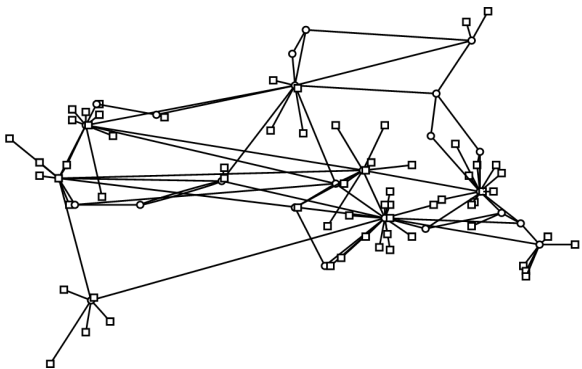

Figure 9. AT&T core topology characterization

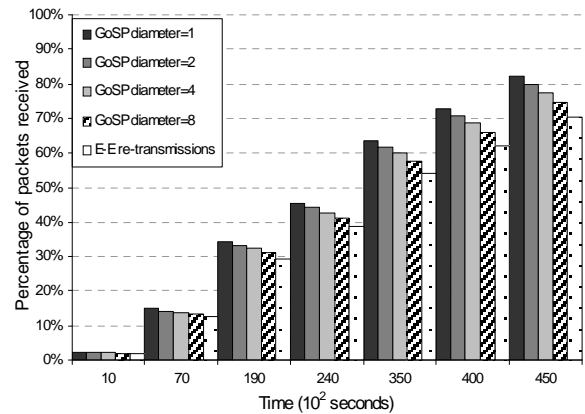

Figure 11. Packets received in sink in GoS re-transmission cases and E-E case at different time samples

Fig. 10 shows a throughput comparative between an E-E case, where lost packets need TCP re-transmissions from the head-end and a GoS case where dropped packets are recovered locally. Due to GoS assigned to the MCN service, 91.04% of discarded packets were recovered with diameter=1, 8.96% with $d$=2 and no packets were re-transmitted with $d$>2. Trend functions are also shown in the chart to allow a performance comparative, with a confidence interval of 12.5Kbps, at a 95% confidence level. Average difference between trend functions is 4.84%.

Fig. 11 shows a comparison between the percentage of packets received at different time samples of a particular flow when dropped packets are E-E recovered by the transport level protocol and when they are re-transmitted locally with $d$=1, $d$=2, $d$=4 and $d$=8 diameters. For instance, at 35000s, 55.79% of E-E traffic has been received; at the lowest GoS level case ($d$=8), 58.12% of packets have already been received, in the $d$=4 case, 60.04% of packets, in the $d$=2 case 61.83% of packets and in the best GoS level case, when $d$=1, 62.91% of packets have been received.

Therefore, the more GoS capable nodes crossed by the LSP, the higher the probability for local re-transmissions with optimal diameter=1. Hence a MPLS service provider would assign flows with the highest GoS level to an LSP that crosses more GoS nodes.

Fig. 12 shows a packet loss comparative between a no GoS case, where a lost packet need a TCP re-transmission from the head-end and a GoS case where discarded packets can be recovered locally; therefore, these would not be considered as lost packets at the head-end. Trend functions are also shown, with a confidence interval of 0.21%, at a 95% confidence level and an average difference between trend functions of 1.32%.

This way, we conclude that a significant part of discarded traffic will not have to be recovered end-to-end by transport layer protocol due to GoS local re-transmissions. Furthermore, including GoS capable nodes in bottlenecks we obtain an improvement in the number of packets delivered for MCN services in the Internet, with a better use of network resources.

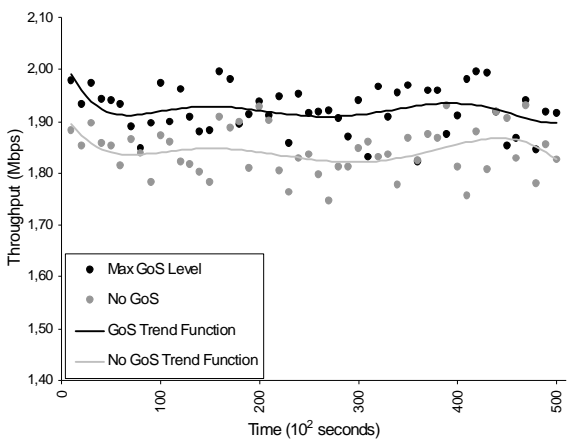

Figure 10. Throughput sampling comparative between GoS and E-E re-transmissions

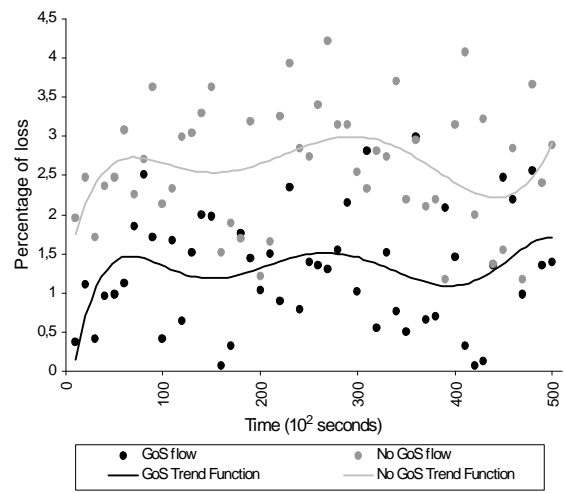

Figure 12. Percentage of packet loss of GoS and E-E flows

## VI. CONCLUDING REMARKS

This article discusses GoS as a local traffic recovery technique in a MPLS domain with the aim of improving the network performance for MCN services in the face of congestion. We have first defined and discussed the requirements for GoS over MPLS. Then, we have explained that GoS signalling for MCN services with requirements of low delay and high reliability is possible. The scalability of the proposal has been analytically studied and, finally, the benefits due to local re-transmissions of discarded traffic with respect to end to end re-transmissions have been evaluated. Further work should include the evaluation and comparison of different network scenarios under different real traffic distributions.

AUTHORS PROFILE

Fco. Javier Rodríguez-Pérez received his Engineering degree in Computer Science Engineering at the University of Extremadura (Spain) in 2000, where he is currently a professor and a Ph. D candidate of GITACA group. His research is mainly focussed on QoS and traffic engineering, packet classification and signalling development over IP/MPLS systems.

José-Luis González-Sánchez is a full time associate professor of the Computing Systems and Telematics Engineering department at the University of Extremadura, Spain. He received his Engineering degree in Computer Science and his Ph.D degree in Computer Science (2001) at the Polytechnic University of Cataluña, Barcelona, Spain. He has worked, for years, at several private enterprises and public organizations, accomplishing functions of System and Network Manager. He is the main researcher of the Advanced and Applied Communications Engineering Research Group (GÍTACA) of the University of Extremadura. He has published many articles, books and research projects related to computing and networking.

Alfonso Gazo-Cervero received his PhD in computer science and communications from the University of Extremadura. He is currently a member of the research and teaching staff as assistant proffesor in GITACA group. His research interests are related mainly to QoS provision over heterogeneous networks, capacity planning, routing protocols and overlay networks.